# Measurement and control of the fast ion redistribution on MAST

M. Turnyanskiy[1], C. D. Challis[1], R. J. Akers[1], M. Cecconello[2], D. L. Keeling[1], A. Kirk[1], R. Lake, S. D. Pinches[1,3], S. Sangaroon[2], I. Wodniak[2]

[1] EURATOM/CCFE Fusion Association, Culham Science Centre, Abingdon, Oxon, OX14 3DB, UK
[2] Dept. of Physics and Astronomy, Uppsala University, Sweden (EURATOM-VR Association)
[3] now at : ITER Organization, POP, Science Division, Route de Vinon sur Verdon, 13115 St. Paul-lez-Durance, France

E-mail: mikhail.turnyanskiy@ccfe.ac.uk

**Abstract**. Previous experiments on MAST and other tokamaks have indicated that the level of fast ion redistribution can exceed that expected from classical diffusion and that this level increases with beam power. In this paper we present a quantification of this effect in MAST plasmas using a recently commissioned scanning neutron camera. The observed fast ion diffusivity correlates with the amplitude of n=1 energetic particle modes, indicating that they are the probable cause of the non-classical fast ion diffusion in MAST. Finally, it will be shown that broadening the fast ion pressure profile by the application of neutral beam injection at an off-axis location can mitigate the growth of these modes and result in the classical fast ion behaviour.

## 1. Introduction

A key tool to control the q-profile and provide heating and non-inductive current drive in tokamak plasmas is neutral beam injection, which relies on good fast ion confinement. Driving current by the injection of fast ions is particularly important in spherical tokamaks due to the limited applicability of other non-inductive current drive techniques and because of the limited space available for neutron shielding of a solenoid. Plasma operation with high neutral beam power and the associated large fast ion pressure poses the risk of destabilising low-n fast ion driven instabilities, which have the potential to degrade the fast ion confinement and lead to a reduction in core heating and current drive efficiency. The ability to measure the radial distribution of fast ions in a tokamak plasma is a prerequisite for the identification of conditions where good fast ion confinement can be achieved. In highly driven plasmas, as is the case for present tokamaks and concepts such as the Component Test Facility (CTF) [1], Neutral Beam Injections (NBI) can provide fast ion pressure profile control as well as plasma heating and non-inductive current drive. It is necessary, therefore, to optimise the use of NBI systems to fulfil these multiple functions so as to simultaneously achieve long pulse or steady-state operation while providing plasma performance control and avoiding deleterious plasma instabilities.

For NBI heating and current drive to be effective the fast ions must remain confined in the plasma until they are thermalised. Some plasma MHD instabilities can disturb the orbits of the fast ions, causing them to move radially outwards from the plasma core and, in some cases, leave the plasma altogether. A prime candidate is a class of instabilities that are destabilised by the fast ions themselves. In this case the injected beam ions generate high pressure in the core of the plasma, which destabilises modes that transport the fast ions away from the plasma centre. One of the most commonly reported causes responsible for fast ion confinement degradation is n = 1 beam driven fishbone activity [2]. For example, in work reported previously on MAST [3-5] it was necessary to enhance the fast ion diffusivity above classical values by ~1m²/s to simulate the experimentally observed neutron rate and plasma stored energy at high NBI power. It is thought that this 'anomaly' was associated with the observed n = 1 fishbone MHD mode. This mode is localised in the plasma core region, where almost all the beam ions in MAST are in passing orbits and can thus contribute to current drive. Beam ion losses caused by the mode could therefore have an impact on current drive efficiency.

In this paper we present the results of new experiments that have been made possible by recent MAST system improvements. Firstly, the upgraded MAST NBI system [6] now comprises two identical beam sources which allow the heating power level to be varied by a factor of two while maintaining the same beam deposition profile and initial fast ion energy distribution. Secondly, a scanning neutron camera has been installed and commissioned [7]. This has four collimated lines-of-sight that can be scanned across the plasma between plasma pulses. The MAST neutron emissivity is dominated by beam-plasma interactions, partly due to the large energy difference between the beam energy and the plasma ion temperature (~1-2keV), and consequently the radial distribution of the neutron emission is closely coupled to the NBI fast ion profile. This method is particularly sensitive to ions with energies close to the beam injection energy, which are estimated to generate more than 50% of NBI driven current in MAST. Thus these two features of the present MAST capability provide an ideal opportunity for the quantification of anomalous redistribution of high energy fast ions and the optimisation of plasma conditions to allow efficient NBI heating and current drive.

## 2. Tokamak systems, diagnostics and modeling codes

MAST plasmas have typical parameters of $R \sim 0.85m$, $a \sim 0.65m$, $I_p < 1.3MA$, $B_t < 0.5T$ and the tokamak is equipped with a comprehensive set of plasma diagnostics that enable interpretative modelling of the thermal plasma and fast ion distribution. For the experiments reported here the two co-injected deuterium NBI sources were operated at 60kV, each producing 1.5MW of heating power. The NBI power was split between the main, half and third energy components in the ratio 89%:8%:3%, respectively, and both beams were injected along the equatorial plane of the tokamak with a tangency radius, $R_{tan}$, of 0.7m.

The neutron camera measures the spatially collimated, line integrated, D-D 2.5MeV neutron particle flux emitted by the plasma. The data are recorded with a spatial resolution of ~5cm and a time resolution of up to 1ms (typically 5-10ms for a representative MAST discharge to achieve 10-15% statistical measurement error). The camera has a flexible and modular design for both gamma-ray and neutron shielding and collimation, and the detectors view the plasma through a thin stainless steel flange (3 mm) to avoid attenuation of the neutron flux. A set of liquid scintillators coupled to a fast digitizer is used for neutron/gamma-ray digital pulse shape discrimination [7]. Two lines-of-sight are orientated to view radially in the equatorial plane of the tokamak, at two different tangency radii. The other two viewing lines are vertically displaced and angled downwards at the same two tangency radii. In the work reported here, only data from the equatorial pair of detectors has been used. The camera has been installed on a spatial scanning system which is capable of varying the tangency radii of the viewing lines in the range $R_{NC} \sim 0m$ to $R_{NC} \sim 1.21m$, covering the entire core region of the MAST plasma as shown in FIG.1. Exploiting the shot-to-shot reproducibility of MAST plasmas, the data can be combined from a series of identical plasma pulses where different camera locations were used to reconstruct the full radial profile of the neutron emissivity.

Accurate interpretative simulation of MAST plasmas is possible due to a comprehensive set of diagnostics. The electron density and temperature are measured using a 160 point, 200Hz Thomson scattering diagnostic system and a single time 300 point system, (radial resolution, $\Delta r \sim 5mm$) [8]. The edge neutral density and edge density gradients are reconstructed from an absolutely calibrated linear $D_\alpha$ camera [9][10] combined with local electron density and temperature from the Thomson scattering system. The ion temperature and rotation profiles are measured using charge exchange recombination spectroscopy. The performance of the beams is monitored by beam emission imaging spectroscopy [11], delivering time-resolved measurements of the beam fast neutral density and reinforcing data on the NBI geometry for beam modelling codes. A 2D visible bremsstrahlung imaging camera [12], which provides $Z_{eff}$ profiles using Thomson scattering data, and an infrared camera for diagnosing power flow to the diverter targets and particle loss to in–vessel components [13] are also available. The direct measurements of the current profile are taken by a 35 point Motional Stark Effect (MSE) diagnostic [14]. The total neutron rate is measured by an absolutely calibrated 1.34 g $^{235}$U fission chamber (FC) [15] located in close proximity to the vessel with a time resolution of 10µs.

The Larmor orbit corrected Monte Carlo TRANSP code [16] has been used to model the thermal and fast ion components of the plasma. In this approach at each time step, a "random" gyro-phase is selected then a Newton Raphson search algorithm is used to find the position of a particle point in real space that corresponds to that guiding centre. The local particle plasma parameters are then used in the collision operator. The effect over many poloidal orbits and many Monte Carlo particles then is to create a distribution function that is close to that which would be created by a full gyro-orbit calculations. The simulations incorporate the measured profiles of $T_e$, $n_e$, $T_i$ and $Z_{eff}$ (with carbon as the dominant impurity) inside the separatrix. $T_e$, $n_e$ in the scrape-off layer are taken from Langmuir probe measurements and the neutral particle fluxes from the $D_\alpha$ camera. The experimentally measured radial profiles of the toroidal rotation are also incorporated into simulations. The plasma boundary is provided from MSE constrained EFIT equilibrium reconstructions [17]. TRANSP can provide a synthetic neutron camera diagnostic based on an assumed fast ion and plasma model and the geometry of the MAST camera, allowing the raw measurement data to be compared for consistency with simulations without the need for profile reconstruction from the array of line integral measurements.

## 3. Fast ion redistribution with on-axis NBI

MAST plasmas have been specifically developed to quantify the effect of fast ion redistribution at different NBI power levels. To isolate the effect of n=1 energetic particle modes, it was first necessary to exclude other MHD phenomena that are known to produce fast ion redistribution, such as q=1 sawteeth and 'long lived mode' inabilities [18]. The abrupt reconnection events at sawteeth introduce temporal variations in the thermal plasma parameters as well as potentially disturbing the fast ion distribution, and continuous n=1 MHD modes, such as the 'long lived mode', have been seen to produce significant fast ion redistribution [19]. For this reason plasma scenarios were used that included a large plasma volume during the current ramp-up phase, coupled with a fast current ramp rate. This technique is commonly used in tokamaks to arrest the initial current penetration to the plasma centre and produce a plasma at the start of the current flat-top with a minimum value of q above unity. In MAST this provided a period early in the current flat-top where q=1 sawteeth and 'long lived modes' were absent. In addition, it was found that avoiding the transition to H-mode was advantageous to avoid a significant temporal evolution of plasma parameters due to the initial L-H transition and subsequent ELM instabilities. In the experiments reported here, the plasma was maintained in L-mode by adjustment of the fuelling and magnetic configuration. Thus plasma scenarios were developed to maximise the shot-to-shot reproducibility, necessary to reconstruct the neutron profile from a series of similar plasma pulses, and to minimise the disturbance due to other MHD phenomena.

The radial neutron emission profile has been measured for on-axis NBI using an 800kA L-mode plasma in a Double Null Divertor (DND) magnetic configuration. The central plasma density was ~4×10$^{19}$ m$^{-3}$ and the magnetic field was ~0.45T. The first set of plasmas were heated with a single NBI beam, scanning the neutron camera location between pulses. To test the similarity of the two beams, single beam reference cases were made with each of the MAST sources. The neutron rates in these two cases differed by only ~3-5%, which is within the uncertainty due to small variations in NBI parameters and confirms the validity of the experimental approach. Then a similar set of plasmas were heated using both beams, but maintaining the same plasma density as in the single beam cases. This provided a two-point power scan for the assessment of the effect of increasing the beam power on the resulting fast ion distribution.

The time evolution of typical plasma parameters is shown in FIG.2 for representative pulses with one (red curves) and two (black curves) beams, respectively. The digital plasma current and density feedback control allowed the ~1.5MW and ~3MW beam heated plasmas to be controlled with very similar total current and density profiles. The addition of the second NBI source at the same starting time of ~0.15s is marked by an increase in the observed total neutron rate, $R_n$, compared with the single beam pulse, as shown in FIG. 2(left). The increase in central electron and ion temperatures and toroidal rotation was only limited in the range 10-25% with $n_e$ and $T_e$ profiles at t=0.27s shown in FIG.2 (right).

Data from a mid-plane Mirnov coil are presented in FIG.3 and shows extensive n=1 fishbone activity between 50 and 100 kHz in plasmas with the same plasma density and boundary location. The signals show that the MHD has a similar character at the two different power levels (1.5MW and 3MW), but the amplitude of the MHD signal has more than doubled in the higher NBI power case (FIG.3, left). The onset of the 'long lived mode' at 0.28s in both pulses significantly degrades the fast ion confinement as well as other plasma properties [19]. The phase with the 'long lived mode' is ignored (t>0.27s) in the analysis reported here. A correlation of the central neutron camera channel (R=0.9m) signal with fishbone bursts in the 3MW powered discharge is shown in FIG.3 (right), suggesting that radial transport of the fast ions is affected by this beam-driven MHD activity. Statistical correlation plots of the amplitude of the observed n=1 fishbone activity and the change in both relative and absolute neutron rates measured by the fission chamber diagnostic [15] is presented in FIG.4 for a set of similar MAST NBI-heated discharges. Some statistical data variation can be attributed to the small variations in the plasma shape and radial location in the database used to generate FIG.4. Nevertheless a clear correlation is apparent, with some fishbone bursts leading to up to 20% loss in the neutron rate, providing additional evidence that these MHD events are responsible for the associated redistribution of the fast ions.

## 4. TRANSP modelling of the fast ion distribution

The fast ion transport model used in the TRANSP code allows the introduction of an *ad hoc*, pitch angle-independent, beam-ion diffusion coefficient, $D_{an}$, to be added to the classical diffusion model to approximate the effect of an anomalous fast ion redistribution of the sort observed in these experiments. While this *ad hoc* TRANSP facility is extremely helpful in "quantifying" and comparing the degree of the fast ion redistribution divergence from the classical fast ion behaviour, it by no means proves the diffusion nature of processes involved. Theoretical investigations aiming to understand and describe the exact mechanism for evolution of the fast ion function in the presence of n=1 mode is on-going at the moment but are outside of the remit of this experimental work. Here we are concentrating on the optimisation of plasma conditions to allow efficient NBI heating and current drive by avoiding this anomalous fast ion redistribution and studying its power scaling dependence. The *ad hoc* TRANSP model also allows the magnitude of this anomalous fast ion diffusion to be varied as a function of time, space and energy of the fast particles. In the simulation presented in this paper the effective anomalous fast ion diffusivity is assumed to be constant in space, time and energy.

Both experimentally-measured and TRANSP-simulated neutron rates and plasma stored energies calculated by the EFIT code [17] for plasmas with injected power ~1.5MW and ~3MW are shown in FIG.5. Up to the onset of the "long lived mode" (~0.27s), which is known to degrade the fast ion confinement in MAST, the plasma discharges with $P_{NBI}$ =1.5 MW on MAST usually exhibit only low intensity fast particle-driven MHD. In this case TRANSP modelling assuming classical beam deposition and using the Chang-Hinton model [20][21], agrees well with the experimentally-measured total neutron rate and plasma stored energy (see FIG.5. left). The agreement of the modelling with experiment indicates that the beam ion population in the low power ($P_{NBI}$ < 2 MW) co-injected MAST plasmas studied here evolves predominately due to Coulomb collisions and charge exchange phenomena and confirms the classical behaviour of the fast ions. The fast ion losses were low (a few %) and dominated by charge exchange [22] [23]. There was no strong evidence of any anomalous slowing-down, for example due to fast particle MHD interactions. Comparison with the experimentally measured neutron rate in higher power (3MW) discharges indicates that a diffusion coefficient of roughly $D_{an}$~2m$^2$/s is required to account for the measurements, comparable to that previously reported from DIII-D [24], AUG [25], JET [26] and NSTX [27]. An assumed level ($D_{an}$~2 m$^2$/s) of the anomalous fast ion redistribution also improves the agreement with the stored energy measurements and provides a useful check on the hypothesis of enhanced fast ion transport, further supporting the conclusion drawn from the analysis of the neutron rate measurements. Other plasma parameters with large systematic uncertainties that may affect the fast ion simulation, such as neutral density or edge ion temperature, were varied within realistic upper and lower limits. The simulated neutron rate and stored energy have proven to be relatively insensitive to those variations. For example, increasing the neutral density by a factor of ten led to only a very modest (2-3%) drop in the simulated neutron rate and stored energy.

The TRANSP code has also been used to simulate the neutron flux measured by the neutron camera for the various horizontal observation angles used in both sets of pulses. The comparison of the raw measured neutron camera profiles with the synthetic diagnostic provided by the TRANSP simulations is shown in FIG. 6 for the period t=0.25-0.27s in the one and two beam cases. The magnitude of the neutron camera data was cross-calibrated to the total neutron flux obtained by the fission chamber in plasmas with weak MHD activity where no significant anomalous fast ion diffusion was detected. For this set of experiments it was found that the total neutron rate from the TRANSP simulation was in good agreement with the fission chamber measurements for the one beam plasmas and the synthetic neutron profile shape was also consistent with the neutron camera measurements. So this case was used to cross-calibrate the magnitude of the neutron camera signals using the levels generated by the TRANSP synthetic diagnostic. The statistical error bars on the neutron camera data together with estimates of error bars associated with the finite collimator dimensions (1x3cm) are also shown in FIG. 6. Since the measured profile shape presented in FIG. 6 is obtained by scanning the two horizontal neutron camera lines-of-sight across the plasma between pulses, it is not affected by any uncertainties in the cross-calibration method described above. The similarity between the measured and simulated profiles for the one beam case when no significant anomalous fast ion diffusion is assumed is supporting evidence for the hypothesis that relatively little fast ion redistribution occurs in these plasma conditions. Nevertheless, the possibility of some low level of anomalous fast ion redistribution, equivalent to $D_{an} \sim 0.5 m^2/s$, cannot be ruled out even for these low power plasmas within the uncertainties of this analysis.

For the two beam case at the same plasma density, the TRANSP simulation predicts that the neutron emissivity should more than double compared with the one beam plasmas (see FIG. 6). This is due to the factor of two increase in the rate of injection of fast ions and additional contributions from the slightly increased electron temperature (see FIG.2(b)), contributing to longer slowing-down time and hence the steady-state fast ion density, and neutrons generated by beam-beam collisions, which increase as the square of the fast ion density. By comparison the increase in the measured neutron emission for the two beam case is much less than a factor of two compared with the one beam case. The TRANSP simulation, assuming classical fast ion transport, thus significantly overestimates the neutron rate (by ~35-40%), suggesting appreciable anomalous fast ion radial transport at this power level. The data used to generate FIG.6 were time-averaged over a period of 20ms. This means that the measured profiles represent the time-averaged impact of the intermittent bursts of n=1 MHD illustrated in FIG.3. TRANSP simulations assuming redistribution equivalent to an anomalous diffusivity between $2m^2/s$ and $3m^2/s$, comparable to that previously reported in the literature [3] [24-30], show much better agreement with the measurements and are also plotted in FIG.6. These results are further support and provide a consistency check for the results shown in FIG.5, where the experimentally-measured total neutron rate and magnetic plasma stored energy calculated using the EFIT code are compared with TRANSP simulations. The level of assumed anomalous fast ion diffusion required in the simulations to achieve agreement with the fission chamber data and the plasma stored energy is similar to that needed to account for the neutron camera profiles.

TRANSP simulated poloidal projections of the fast ion distributions, integrated over energy and pitch angle, are shown in FIG. 7 for the two simulations that show good agreement with the measurements, i.e. one beam ($P_{NBI} \sim 1.5MW$) with $D_{an} = 0m^2/s$ and two beams ($P_{NBI} \sim 3MW$) with $D_{an} = 2m^2/s$. The tangency point of NBI is highlighted as a cross at R=0.7m, Z=0m. The corresponding fast ion density radial profiles in the plasma mid-plane are also presented in FIG. 7. The effect of the assumed anomalous diffusion on the fast ion confinement in the two beam plasma is clearly visible and results in both a decrease in the central fast ion density, despite the doubling of the NBI power, and a reduction in the peaking of the fast ion density profile.

## 5. Fast ion redistribution with off-axis NBI

The observed increase in fast ion radial transport induced by n=1 energetic ion-driven MHD activity is clearly undesirable, affecting as it does the neutral beam current drive and heating properties. Since energetic particle instabilities are sensitive to the distribution of fast ions, one possible approach for the control of such modes is to broaden the fast ion pressure profile by the application of off-axis neutral beams. At present the MAST NBI systems are fixed in the equatorial plane of the tokamak. However, the flexibility offered by the large MAST vessel has been used to study the effects of off-axis heating by displacing the plasma downwards in a Single Null Divertor (SND) configuration. Experiments to date demonstrate comparable plasma heating effects for off-axis heated discharges to that achieved with on-axis heated discharges with similar plasma current and electron density [3]. Strongly off-axis SND discharges pose a challenge for detailed transport analysis on MAST, due to the majority of diagnostic measurements being located in the vessel mid-plane rather than the equatorial plane of the plasma. In this case some measurements must be extrapolated into the region with normalised poloidal flux $\phi$ in the range $0 < \phi < 0.15$. The central region of MAST plasmas ($0 < \phi < 0.15$) usually exhibits flat $T_e$, $n_e$ and $Z_{eff}$ profiles, allowing the values of these plasma parameters to be estimated at the plasma centre to an accuracy of within ~10-15%. Sensitivity studies, where missing diagnostic data was varied within realistic upper and lower limits, produced consistent results, insensitive to such variations, and resulted in an uncertainty in simulated neutron rate of less than a few percent at most. Modelling assumptions for diagnostic measurements used in SND simulations were further validated by shifting the SND plasma back to the vessel midplane (to optimise diagnostic measurements) in a time scale much shorter (<1ms) than both the energy confinement and beam ion slowing-down times. An example of the TRANSP modelled poloidal projection for the fast ion distribution with off-axis NBI is shown in FIG.8. The off-axis NBI results in much broader deposition of the fast ions when compared to on-axis beam injection. The corresponding TRANSP simulated flux-averaged profiles of the fast ion energy density for similar 800kA plasmas with on- and off-axis NBI are shown in FIG.9 for $D_{an}=0$ m$^2$/s. The corresponding TRANSP modelled neutron radial profiles are also presented and show very similar radial distributions, confirming the validity of the chosen experimental approach. Broadening of the fast ion pressure profile by off-axis NBI reduces the drive for energetic particle modes and fast ion redistribution. Spectrograms from a mid-plane Mirnov coil for two-beam plasmas with 800kA of plasma current and on- and off-axis NBI are compared in FIG.10. The n=1 mode activity is much suppressed and diminished for the off-axis NBI compared with on-axis injection. n=2 and n=3 modes are seen in the off-axis case, but these have a significantly weaker effect in terms of fast ion redistribution. Similar radial profiles of q, measured using the 35 point MSE diagnostic, were observed for on- and off-axis NBI discharges, as shown in FIG.11, further supporting the hypothesis that the change in the fast ion distribution profile is responsible for the change in the energetic particle mode behaviour. The beneficial effect on the fast ion confinement due to the reduction of the n=1 drive has been confirmed by the measured neutron profiles, shown in FIG. 12. In this case, the TRANSP simulations based on classical fast ion transport agree well with the neutron camera profile measurements for both the one beam ($P_{NBI}$~1.5MW) and two beam cases ($P_{NBI}$~3MW). These modelling results are also consistent with the good agreement between TRANSP simulated and experimentally measured total neutron rate and stored plasma energy, suggesting no or very weak (equivalent to $D_{an}<0.5$m$^2$/s) anomalous fast ion radial transport.

An alternative technique to control the growth and mitigate the effect of these instabilities could be to increase the plasma density, thereby reducing the fast ion slowing-down time and hence the fast ion pressure. To make an initial test of this hypothesis preliminary experiments were analysed using the same SND plasma configuration as for the experiments illustrated in FIG.12, but with reduced plasma density (~30% lower). As in the previously described experiments, the power dependence of the fast ion redistribution was monitored by the scaling of neutron emissivity measured by the neutron camera at different power levels. The TRANSP simulated flux averaged profiles of the fast ion energy density and neutron distributions under the assumption of classical fast ion thermalisation ($D_{an}=0$ m$^2$/s) are shown in FIG.9 as dashed lines and can be compared with similar simulations for off-axis NBI plasma with higher density. It can be seen that the plasmas with lower plasma density are predicted to have significantly steeper fast ion pressure gradients, which are found to be sufficient to trigger n=1 energetic particle modes despite off-axis NBI deposition. A spectrogram from a mid-plane magnetic coil for a low density 3MW off-axis NBI discharge is presented in FIG.13 and shows n=1 fishbone activity with similar amplitudes and frequencies to the higher density 3MW on-axis NBI plasma shown in FIG.10 (left). The raw neutron emissivities measured by the neutron camera for the period t=0.25-0.27s are also shown in FIG.13 and indicate strongly non-classical power scaling for 1.5MW and 3MW NBI heating power, where similarly to the simulations shown in FIG.6 and FIG.12 more than doubling of the neutron emissivity is expected, thus confirming the strong anomalous fast ion redistribution in off axis NBI plasmas exhibiting this amplitude of n=1 energetic particle mode activity. These preliminary lower density SND results with off axis NB injection were obtained at lower plasma current ($I_p$=620kA) and, strictly speaking, in presently-available data we cannot separate the effects of density and q-profile variations at this point. Future studies will investigate the effects of varying the density and q-profile more systematically.

### 6. Summary and Discussions

A set of controlled power scaling experiments have been performed to determine the dependence of the MAST plasma fast ion density on NBI heating power. Measured data from a new multi-chord scanning neutron camera, together with measurements of the total neutron rate and plasma stored energy, have been compared with TRANSP simulations with different assumptions about the level of anomalous fast ion diffusivity. These comparisons show that the assumption of classical beam deposition and collisional thermalisation is sufficient to explain measurements at low to moderate NBI power ($P_{NBI}$~1.5MW), although a low level of effective anomalous fast ion diffusion (~0.5m$^2$/s) cannot be ruled out. For high power ($P_{NBI}$~3MW) on-axis NBI heating the classical simulation significantly overestimates the neutron emission, suggesting an appreciable anomalous fast ion radial transport consistent with a diffusion coefficient of roughly $D_{an}$= 2-3m$^2$/s. This is comparable with that previously reported from MAST[3-5], DIII-D[24], AUG[25], JET[26-29] and NSTX[30], leading to a broadening of the fast ion profile, fast ion losses and reduced neutral beam central heating and current drive efficiency.

In these experiments the use of NBI led to the appearance of high intensity n=1 fishbone instabilities. Previously the nonlinear $\delta f$ wave-particle interaction code, HAGIS [31], has been used to describe the evolution of the fast ion distribution function in the presence of $n = 1$ fishbone activity, allowing a quantification of the fishbone-driven fast ion radial diffusion [4-5]. These simulations predicted an MHD-induced radial flux and associated change in the fast ion radial density, leading to a radial broadening of the distribution function and to a higher than classical radial fast ion diffusion associated with this beam ion induced MHD activity. The average rate of diffusion was found to scale with the square of the fishbone amplitude, which is usually a strong function of the injected NBI power. At high NBI power ($P_{NBI}$>3MW) a moderate level of anomalous fast ion diffusion (a few m$^2$/s) was predicted. In the plasmas with low n=1 mode amplitudes ($P_{NBI}$<1.5MW) the simulations predict the fast ion redistribution equivalent to an anomalous fast ion diffusivity that is an order of magnitude smaller (i.e. $D_{an}$<0.3m$^2$/s), making it practically undetectable for experimental measurements and bringing the transport simulations in line with MAST observations. The effect of n = 1 fishbone activity, therefore, provides the most credible mechanism proposed[4-5] so far to explain the experimental results discussed in this paper. Other possible explanations, for which no strong evidence has yet emerged include turbulence-driven redistribution of the fast ions. Experimental investigation of turbulence-driven transport will require the development of plasma scenarios that can avoid the fishbone instabilities that mask the effect of turbulence on the fast ion distribution in present experiments. We have demonstrated that the redistribution of fast ions by fishbone instabilities at high NBI power levels can be mitigated by the application of NBI at an off-axis location. It is believed that this is due to a reduction in fast ion pressure gradients within the plasma and a consequent reduction in drive for the n=1 mode. In high power MAST discharges, mitigating the growth of these modes resulted in a dramatic reduction in the anomalous fast ion redistribution.

Initial results suggest that plasma density control could be an effective tool in controlling the fast ion pressure profiles and mitigating the growth of n=1 fishbone instabilities. However the precise role of and relationship between the density and q-profile variations requires more systematic investigations. Further experiments are planned to study this effect in more details and investigate the potential to access a domain without strong fast ion redistribution with high power on-axis NBI heating at high plasma density.

The investigations reported here highlight the value of neutron profile measurements, not only for diagnosing the fast ion density profile to quantify the effects of instabilities on the fast ion distribution, but also to allow the optimisation of plasma scenarios and even machine design to maximise the effectiveness of heating and current drive systems. Further modelling work is ongoing to describe these fast ion driven modes and their effect on the plasma in more detail. The achieved reduction in anomalous fast ion transport by using off-axis neutral beam injection and controlling the fast ion distribution is predicted to improve NBCD efficiency, and enable the NBI system to be exploited as a potential tool for current profile control.

*This work was funded partly by the RCUK Energy Programme under grant EP/I501045 and the European Communities under the contract of Association between EURATOM and CCFE. The views and opinions expressed herein do not necessarily reflect those of the European Commission.*

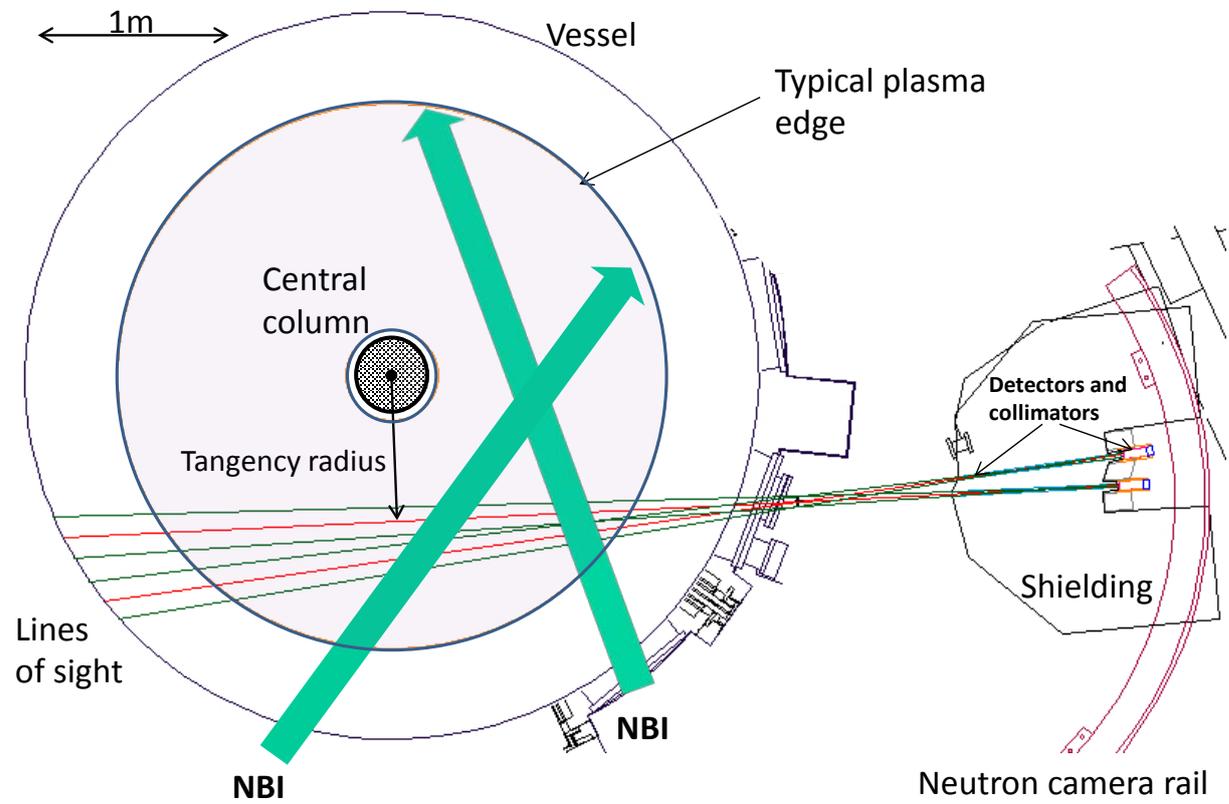

*FIG. 1 The neutron camera has been installed on a spatial scanning system which is capable of varying the tangency radii of the viewing lines in the range $R_{NC} \sim 0m$ to $R_{NC} \sim 1.21m$, covering the entire core region of the MAST plasma.*

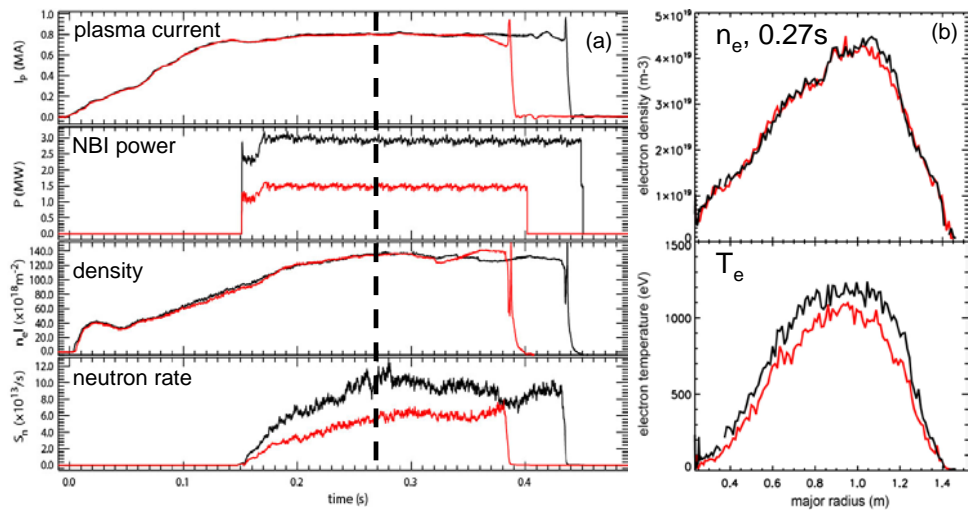

*FIG. 2. (a) Experimental waveforms for two representative plasmas with on-axis NBI power of ~1.5MW (red curves: #26887) and ~3MW (black curves: #26864). (b) $n_e$ and $T_e$ profile measurements from the Thomson scattering diagnostic). The vertical line at t=0.27s in (a) shows the time of the neutron measurements illustrated in Fig. 6 (t=0.25-0.27s) and Thomson scattering measurements (t=0.27s).*

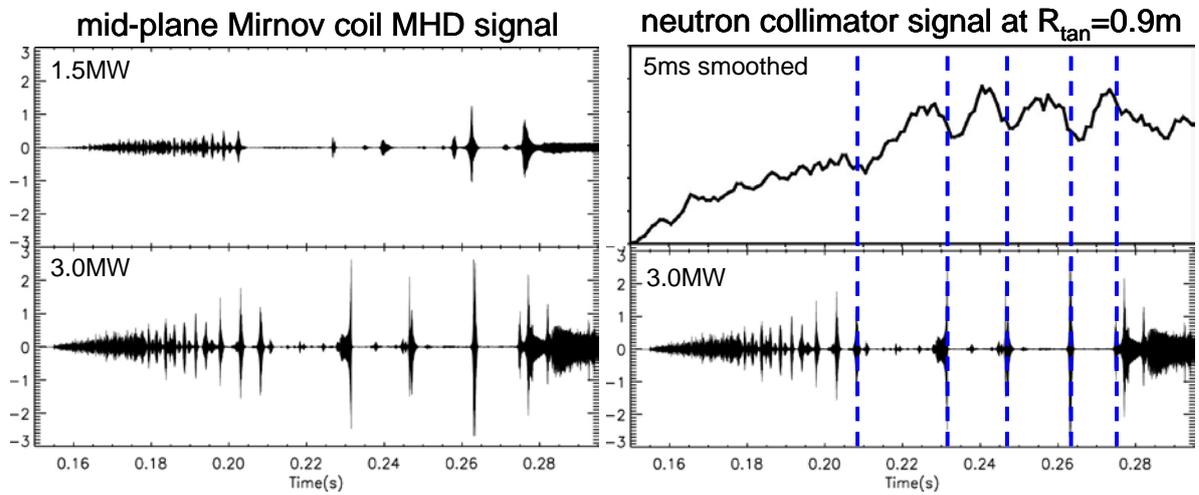

*FIG. 3. Left plot: data from a mid-plane Mirnov coil shows extensive n=1 fishbone activity during NBI heating, more than doubling in amplitude at the higher NBI power (MAST discharges 26887 and 26864). Right plot: correlation of the central neutron camera channel signal with fishbone bursts in 3MW powered discharge ( MAST discharge 26864).*

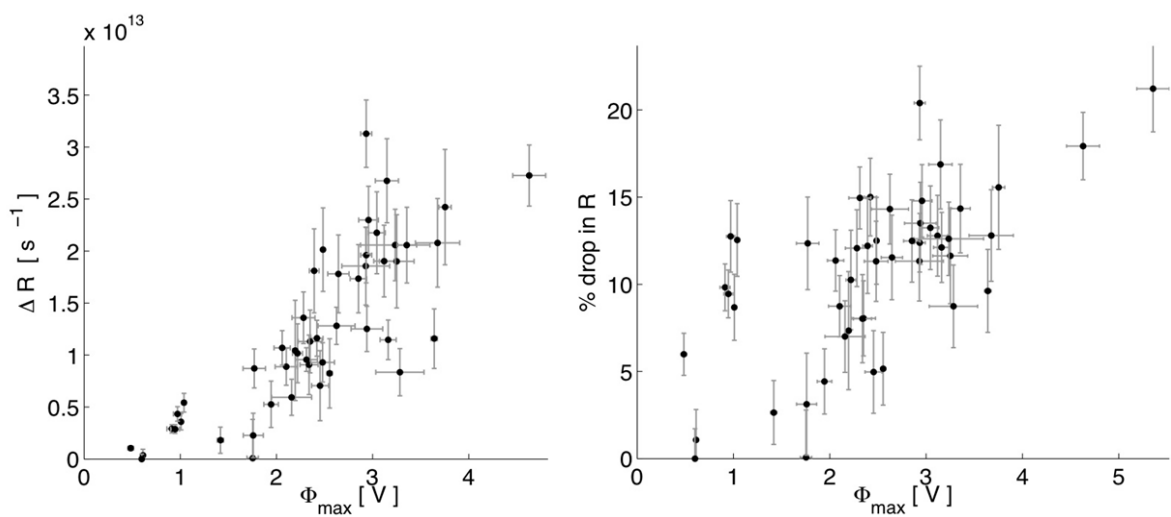

*FIG. 4. Statistical correlation plots for a set of similar MAST NBI heated discharges between the amplitude of the observed n=1 fishbone activity and a change in both absolute (left) and relative (right) neutron rate measured by the fission chamber diagnostic.*

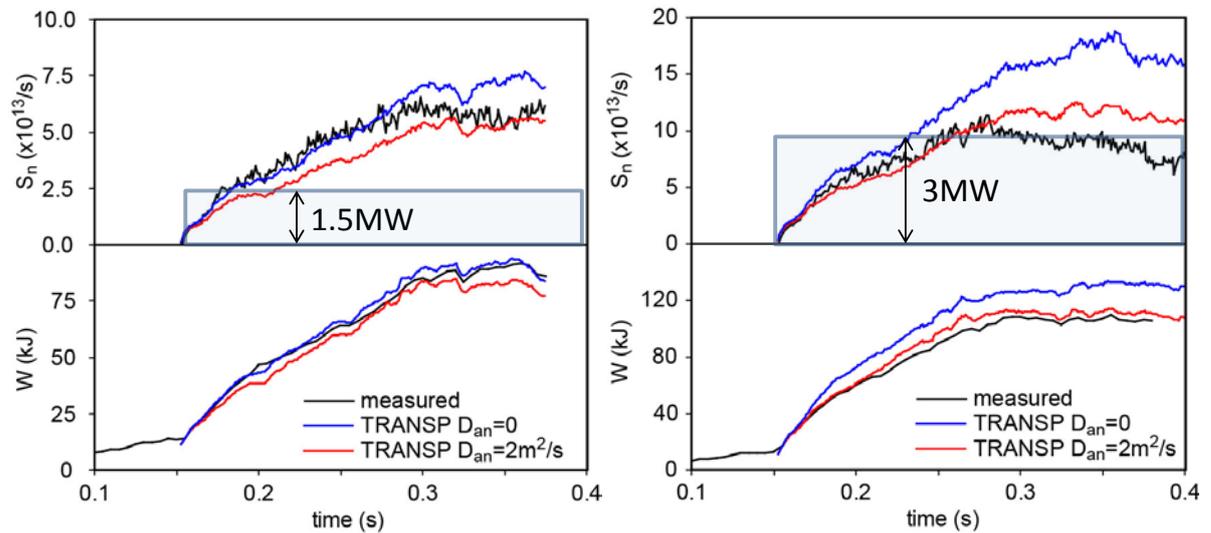

*FIG. 5. Experimentally observed neutron rate and total stored energy measured by EFIT are compared with TRANSP simulations for $D_{an} = 0$ and $2m^2/s$ in 1.5MW (MAST discharge 26887) and 3MW (MAST discharge 26864) NBI heated MAST plasmas. The temporal evolution of the total injected NBI power is also shown.*

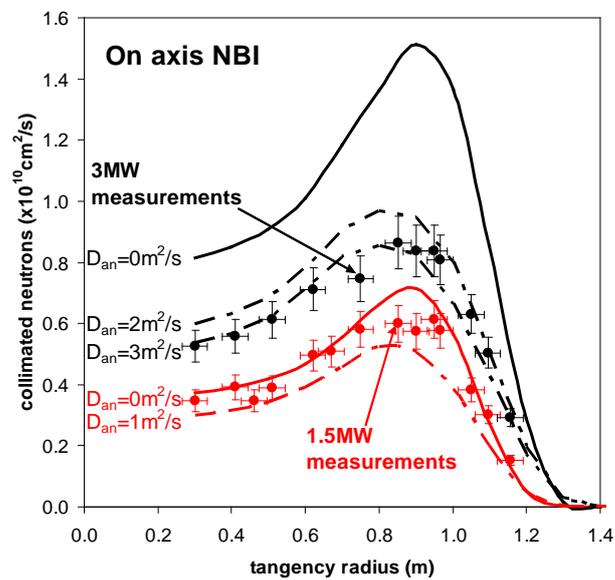

*FIG. 6. TRANSP simulation of the neutron flux measured by the NC for various horizontal observation angles (i.e. various line-of-sight impact parameters) for both sets of discharges studied with $P_{NBI}$ ~1.5MW and 3MW.*

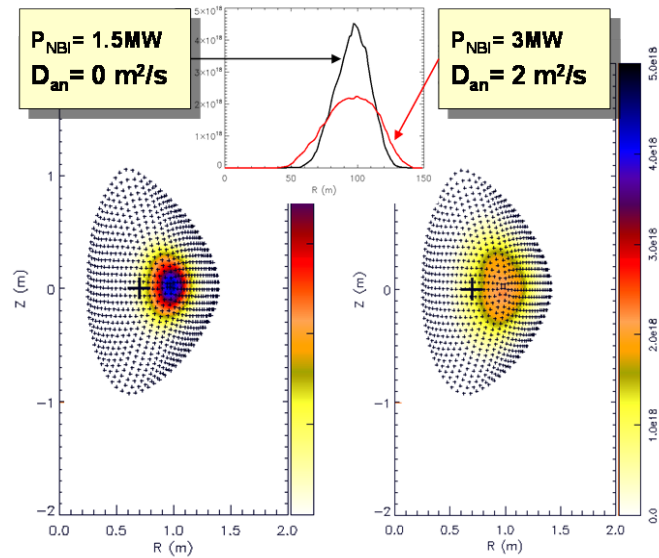

*FIG. 7  TRANSP simulated poloidal projections of fast ion distributions integrated over energy and pitch angle with $D_{an}$=0m²/s at moderate power (MAST #26887, 1.5MW: left) and $D_{an}$= 2m²/s at high NBI power (MAST #26864, 3MW: right).  The midplane fast ion density profiles are also shown in inset.*

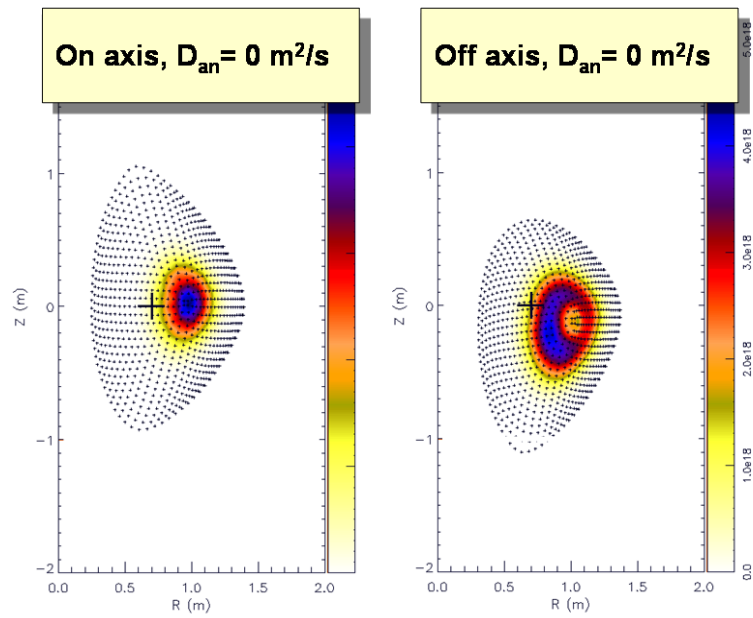

*FIG. 8. TRANSP simulated poloidal projections of fast ion distributions with on- (left) and off-axis (right) NBI, integrated over the energy and pitch angle with $D_{an}= 0m^2/s$ (MAST #26864 and #28283, $P_{NBI}=3MW$).*

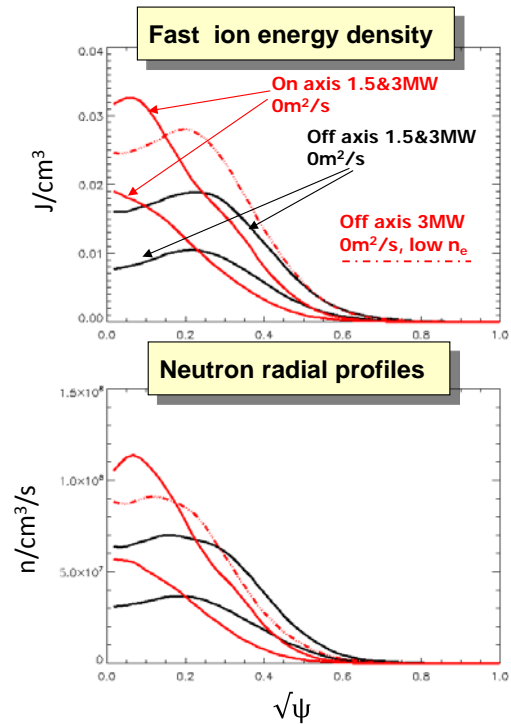

*FIG. 9. TRANSP simulations of the flux averaged distributions of the fast ion energy density and neutrons for on- and off-axis NBI sets of discharges studied with $P_{NBI}$ ~1.5MW and 3MW. The dashed line shows the simulation for a 3MW off-axis discharge with lower line-integrated density ($n_el=1.1 \times 10^{20}$ $m^{-2}$) and $D_{an}=0$ $m^2/s$.*

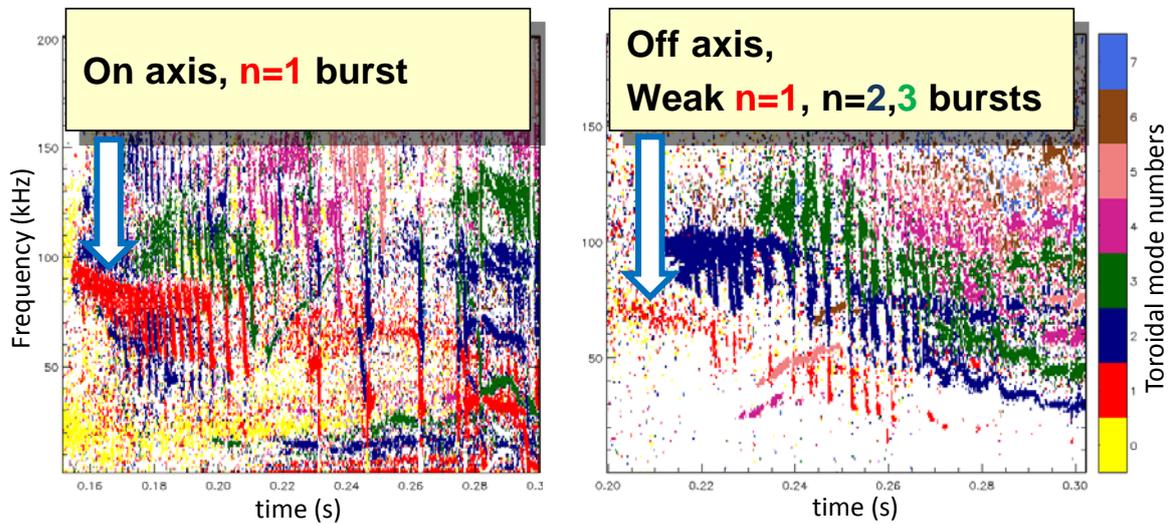

FIG. 10.   Spectrograms from a mid-plane magnetic coil for similar 3MW, 800kA on- (left, MAST 26864) and off-axis (right, MAST 28282) NBI discharges. The observed n=1 mode activity is much weaker for the off-axis NBI compared with on-axis injection.

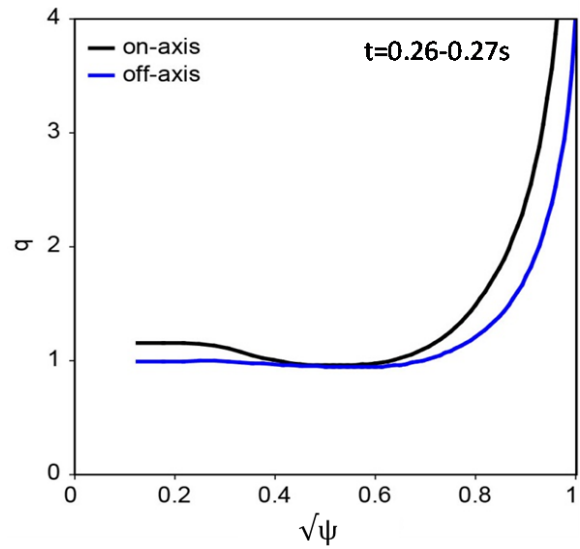

*FIG. 11. Radial q profiles measured by MSE diagnostic for similar 3MW, 800kA on- (black, MAST 26864) and off-axis (blue, MAST 28276) NBI discharges.*

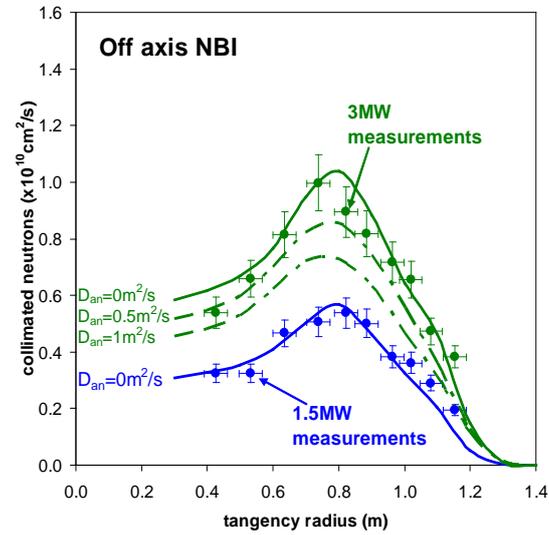

*FIG. 12.TRANSP simulation of the neutron flux measured by the NC for various horizontal observation angles (i.e. various line of sight impact radius) for both sets of off axis NBI discharges studied with $P_{NBI}$ ~1.5MW and 3MW.*

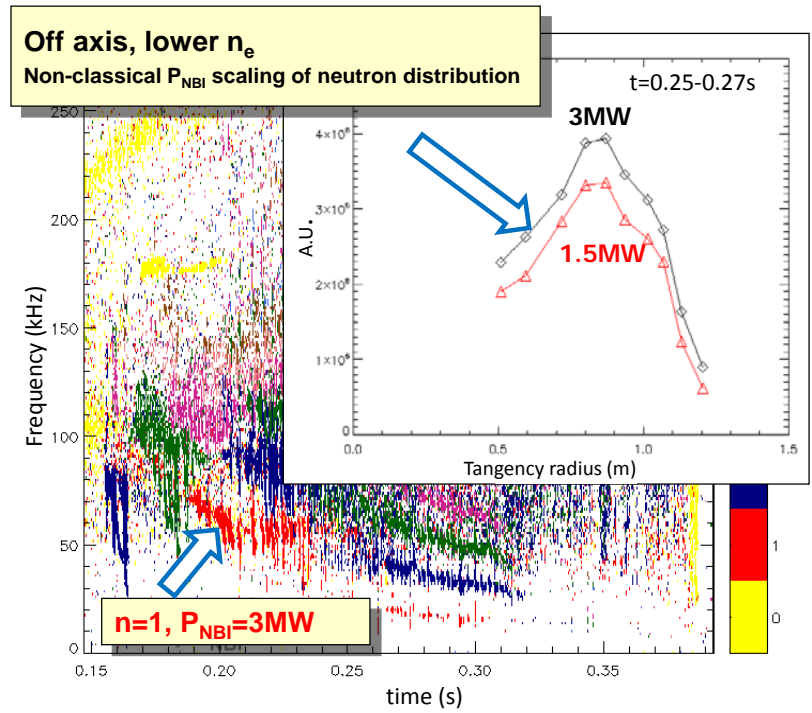

*FIG. 13. Spectrograms from a mid-plane magnetic coil for off-axis 3MW NBI discharges with reduced line averaged plasma density ($n_e l = 1.1 \times 10^{20}$ $m^{-2}$). The raw neutron emissivities measured by the neutron camera for the period t=0.25-0.27s are shown as an inset.*